\documentclass[trackchanges,twocolumn]{aastex7}
\usepackage{graphicx}
\usepackage{natbib}
\usepackage{microtype}
\usepackage{amsmath}
\usepackage{amsfonts}
\usepackage{amsthm}
\usepackage{booktabs}
\usepackage{upgreek}
\usepackage{mathrsfs} 
\usepackage{placeins}
\usepackage{siunitx}

\shortauthors{E.\, Gonzalez et al.}

\begin{document}

\title{Shadowing the Sun I: Testing Synthetic Granulation Spectra with NEID Solar Eclipse Observations}

\author[0009-0002-0987-652X]{Elizabeth Gonzalez}
\affiliation{\PSUAA}
\affiliation{\PSUCEHW}
\email{efg5335@psu.edu}
\author[0000-0002-4677-8796]{Michael L. Palumbo III}
\affiliation{\FlatironCCA}
\email{mpalumbo@flatironinstitute.org}
\author[0000-0001-9596-7983]{Suvrath Mahadevan}
\altaffiliation{Principal Investigator}
\affiliation{\PSUAA}
\affiliation{\PSUCEHW}
\email{suvrath@astro.psu.edu}
\author[0000-0001-6545-639X]{Eric B.\ Ford}
\affiliation{\PSUAA}
\affiliation{\PSUCEHW}
\affiliation{\PSUICS}
\affiliation{\PSUCASt}
\email{ebf11@psu.edu}
\author[0000-0002-9632-9382]{Sarah E.\ Logsdon}
\affiliation{\NOAO}
\email{sarah.logsdon@noirlab.edu}
\author[0000-0003-4384-7220]{Chad F.\ Bender}
\affiliation{\UA}
\email{cbender@arizona.edu}
\author[0000-0002-9082-6337]{Andrea S.J.\ Lin}
\affiliation{\Caltech}
\email{asjlin@caltech.edu}
\author[0000-0001-9626-0613]{Daniel M.\ Krolikowski}
\affiliation{\UA}
\email{krolikowski@arizona.edu}
\author[0009-0005-9178-7017]{Ashutosh Joshi}
\affiliation{\TIFR}
\email{ashutoshjoshi179@gmail.com}
\author[0000-0001-8720-5612]{Joe P.\ Ninan}
\affiliation{\TIFR}
\email{indiajoe@gmail.com}
\author[0000-0001-6160-5888]{Jason T.\ Wright}
\affiliation{\PSUAA}
\affiliation{\PSUCEHW}
\affiliation{\PSETI}
\email{astrowright@gmail.com}
\author[0000-0003-1242-5922]{Ansgar Reiners}
\affiliation{\GU}
\email{Ansgar.Reiners@phys.uni-goettingen.de}
\author[0000-0003-1312-9391]{Samuel Halverson}
\affiliation{\JPL}
\email{samuel.halverson@jpl.nasa.gov}
\author[0000-0003-0241-8956]{Michael W.\ McElwain}
\affiliation{\GoddardESAL}
\email{michael.w.mcelwain@nasa.gov}
\author[0000-0002-2488-7123]{Jayadev Rajagopal}
\affiliation{\NOAO}
\email{jayadev.rajagopal@gmail.com}
\author[0000-0003-0149-9678]{Paul Robertson}
\altaffiliation{Instrument Team Project Scientist}
\affiliation{\UCI}
\email{paul.robertson@uci.edu}
\author[0000-0001-8127-5775]{Arpita Roy}
\affiliation{Astrophysics \& Space Institute, Schmidt Sciences, New York, NY 10011, USA}
\email{arpita308@gmail.com}
\author[0000-0002-4046-987X]{Christian Schwab}
\affiliation{\Macquarie}
\email{mail.chris.schwab@gmail.com}
\author[0000-0001-7409-5688]{Gudmundur Stefansson}
\affiliation{\Schmidt}
\affil{Anton Pannekoek Institute for Astronomy, University of Amsterdam, Science Park 904, 1098 XH Amsterdam, The Netherlands}
\email{g.k.stefansson@uva.nl}
\author[0000-0002-4788-8858]{Ryan C. Terrien}
\affiliation{\Carleton}
\email{rterrien@carleton.edu}

\newcommand{\Schmidt}{Astrophysics \& Space Institute, Schmidt Sciences, New York, NY 10011, USA}
\newcommand{\PSUAA}{Department of Astronomy and Astrophysics, 525 Davey Laboratory, 251 Pollock Road, Penn State, University Park, PA, 16802, USA}
\newcommand{\PSUCEHW}{Center for Exoplanets and Habitable Worlds, 525 Davey Laboratory, 251 Pollock Road, Penn State, University Park, PA, 16802, USA}
\newcommand{\PSUARC}{Astrobiology Research Center, 525 Davey Laboratory, 251 Pollock Road, Penn State, University Park, PA, 16802, USA}
\newcommand{\PSETI}{Penn State Extraterrestrial Intelligence Center, 525 Davey Laboratory, 251 Pollock Road, Penn State, University Park, PA, 16802, USA}
\newcommand{\FlatironCCA}{Center for Computational Astrophysics, Flatiron Institute, 162 Fifth Avenue, New York, NY 10010, USA}
\newcommand{\PSUICS}{Institute for Computational and Data Sciences, Penn State, University Park, PA, 16802, USA}
\newcommand{\PSUCASt}{Center for Astrostatistics \& Astroinformatics, 525 Davey Laboratory, 251 Pollock Road, Penn State, University Park, PA, 16802, USA}
\newcommand{\UA}{Steward Observatory, University of Arizona, 933 N.\ Cherry Ave, Tucson, AZ 85721, USA}
\newcommand{\Caltech}{Department of Astronomy, California Institute of Technology, 1200 E California Blvd, Pasadena, CA 91125, USA}
\newcommand{\JPL}{Jet Propulsion Laboratory, California Institute of Technology, 4800 Oak Grove Drive, Pasadena, California 91109}
\newcommand{\NOAO}{U.S. National Science Foundation National Optical-Infrared Astronomy Research Laboratory, 950 N.\ Cherry Ave., Tucson, AZ 85719, USA}
\newcommand{\GoddardESAL}{Exoplanets and Stellar Astrophysics Laboratory, NASA Goddard Space Flight Center, Greenbelt, MD 20771, USA}
\newcommand{\TIFR}{Department of Astronomy and Astrophysics, Tata Institute of Fundamental Research, Homi Bhabha Road, Colaba, Mumbai 400005, India}
\newcommand{\UCI}{Department of Physics \& Astronomy, The University of California, Irvine, Irvine, CA 92697, USA}
\newcommand{\Macquarie}{School of Mathematical and Physical Sciences, Macquarie University, Balaclava Road, North Ryde, NSW 2109, Australia}
\newcommand{\UAm}{Anton Pannekoek Institute for Astronomy, 904 Science Park, University of Amsterdam, Amsterdam, 1098 XH}
\newcommand{\Carleton}{Carleton College, One North College St., Northfield, MN 55057, USA}
\newcommand{\GU}{Institut für Astrophysik und Geophysik, Friedrich-Hund-Platz 1, 37077 Göttingen, Germany}

\begin{abstract}
The subtle Doppler radial velocity (RV) signature of an Earth-mass exoplanet in the Habitable Zone around a nearby Sun-like star is difficult to detect due to stellar variability. While multiple techniques aim to mitigate rotationally-modulated stellar activity, granulation, introducing 0.1-1 m s$^{-1}$ RV shifts, currently sets the floor for detecting Earth-mass RV signals for low-activity stars. Observations of the Rossiter-McLaughlin (RM) effect during a solar eclipse, with a peak-to-trough amplitude of $\sim$1.4 km s$^{-1}$, are a rigorous test of granulation and convective blueshift in the solar spectrum. Using NEID solar feed data of the October 2023 solar eclipse and \texttt{GRASS-E} (a geometry-modified version of \texttt{GRASS}), we compare the observed RVs for 22 lines with the simulated RVs from the synthetic granulation-driven spectra. Residuals of the \texttt{GRASS-E} Fe I 5434 $\text{\AA}$ model show an RMS of 13.47 m s$^{-1}$ over the RM signal. We find that a model with an optimized convective blueshift polynomial for Fe I 5434 $\text{\AA}$ significantly outperforms \texttt{GRASS-E} with an RMS of only 2.90 m s$^{-1}$, suggesting that \texttt{GRASS-E} is currently limited by sparse center-to-limb observations. By combining a set of spectral lines, we achieve a weighted RMS of 5.2 m s$^{-1}$ over the RM signal ($\sim$0.37\%), a level of accuracy well beyond the reach of current exoplanet transit observations.
In all, we demonstrate that solar eclipse observations allow us to recover the center-to-limb variation of the convective blueshift for individual spectral lines. 
\end{abstract} 

\section{Introduction}
\label{sec:Introduction}
The first exoplanet orbiting a main-sequence star was discovered in October 1995 using the radial velocity (RV) method \citep{1995Natur}. Since this discovery, the RV method remains a crucial technique for confirming and characterizing candidate exoplanets, and is the only current method that can potentially measure the masses of terrestrial exoplanets orbiting within the Habitable Zone of nearby stars, some of which will be observed by the planned Habitable Worlds Observatory (HWO, \citealt{2026arXiv260211046A}). Advancements in present-day spectrographs, such as optical fiber double scramblers \citep{2015ApJ...806...61H}, stable environment control systems \citep{2019JATIS...5a5003R}, and laser frequency comb-based wavelength calibration \citep{2008Natur.452..610L}, allow for sub-m s$^{-1}$ RV precision. However, an Earth-mass exoplanet orbiting within the Habitable Zone of a solar-mass star produces a Doppler shift of only 8.9 cm s$^{-1}$, which currently remains beyond our reach, not primarily due to instrumentation, but also due to astrophysical stellar variability. Several strategies have been designed to model and mitigate the various forms of stellar variability, which occur on many different timescales. For example, pressure modes (acoustic modes excited by near-surface convection) can be reduced by binning over the pressure mode period \citep{Chaplin2019}. Rotationally-modulated magnetic activity, such as dark spots, bright plages, and faculae, can be efficiently modeled via Gaussian processes (e.g., \citealt{Rajpaul2017,Haywood2017,Kosiarek2020, 2025AJ....169...92B}). In addition, several statistical methods, such as \texttt{StarSim} \citep{2016A&A...586A.131H}, \texttt{starry} \citep{2019AJ....157...64L}, and \texttt{SOAPv4} \citep{2025A&A...702A..84C}, simulate the photometric and spectroscopic impacts of magnetically active regions. However, should such sources of stellar activity be successfully removed, granulation would still hinder the detection of Earth-mass exoplanets \citep{2023A&A...676A..82M, 2023AJ....165...98L}.

Granulation, a convective phenomenon, is the interplay between rising hot plasma called granules and cool intergranular lanes of falling plasma. Granulation can be further decomposed based on magnetic field strength, with magnetohydrodynamic (MHD) studies finding distinct contributions from granules, magnetic and non-magnetic intergranular lanes, and magnetic bright points \citep{2013ApJ...763...95C}. Granules on solar-type stars have lifetimes of approximately 10 minutes and range in size from 1300 to 10000 km \citep{Gray2008}. Granulation RV variability depends on line formation temperature \citep{2022A&A...664A..34A, 2024A&A...683A.106A}, wavelength region \citep{2023A&A...680A..62E}, and ionization and excitation potential \citep{2025arXiv250909824S}. 
The collective motion of granules and integranular lanes integrated over the solar disk results in 0.1-1.0 m s$^{-1}$ RV shifts \citep{Rincon2018}. Therefore if left uncorrected, granulation can easily overshadow the Doppler signatures of Earth-mass planets, leaving these exoplanets undetectable. 

The effects of granulation in perturbing the position and shape of spectral lines have been widely discussed in the literature. Following the nomenclature of \cite{Gray2008}, these perturbations are often referred to as first, second, and third signatures. The first signature refers to the supra-thermal broadening of spectral lines that makes the lines wider and shallower. The second signature refers to the asymmetries of spectral lines, and the third signature refers to the net convective blueshift of spectral lines. The latter two signatures are a direct result of the granules being hotter, larger and therefore contributing more light than the intergranular lanes \citep{Dravins}. Center-to-limb variations (CLV) have been measured in disk-resolved solar observations (e.g., \citealt{2018A&A...611A...4L, 2019A&A...622A..34S, 2019A&A...624A..57L}). At disk center, spectral line bisectors manifest in a ``C'' shape, whereas towards the solar limb they resemble more of a ``$\backslash$'' shape. In addition, the net convective blueshift declines towards the limb, with some lines demonstrating slight redshifts near-limb \citep{1907AN....173..273H, 2023A&A...680A..62E}. These CLVs are a direct result of different viewing angles across the solar disk. At disk center, granules are viewed deeper into the photosphere, whereas towards the limb the optical path increases so granules are viewed higher in the photosphere. As a result, different temperature contrasts, velocity flows, and therefore line profiles are observed across the solar disk.

It has been claimed that the RV effects of granulation can be reduced via observation strategies such as averaging multiple observations spaced by hours in a single night \citep{2011A&A...525A.140D, 2015A&A...583A.118M}. In addition, 3D MHD simulations combined with radiative transfer have been used to quantify granulation-driven line perturbations \citep{Cegla2013, Cegla2018, Cegla2019a, 2025MNRAS.539.2248F}. A strong correlation has been found between spectral line bisector statistics and the granulation-induced RV shift \citep{Cegla2019a}, suggesting that precise measurements of line bisectors could trace, and therefore help mitigate, granulation-driven RV shifts. However, such MHD simulations are based on a single simulated line, whereas real stellar spectra have thousands of lines. To address this limitation, \cite{2022AJ....163...11P, 2024AJ....168...46P} developed the GRanulation And Spectrum Simulator (\texttt{GRASS}, available on Zenodo, \cite{palumbo_iii_2026_23814656}) that creates synthetic spectra with time resolved granulation-driven variability. Unlike the MHD simulations, \texttt{GRASS} uses high spectral resolution disk-resolved Laser Absolute Reference Spectrograph (LARS, \citealt{2017A&A...607A..12L}) observations of the Sun to create synthetic disk-integrated spectra. \cite{2024AJ....168...46P} found that use of bisector statistics can only remove 25-35\% of granulation noise. Hence, detection of Earth-mass exoplanets requires accurately accounting for granulation-driven RV variability. \texttt{GRASS} successfully reproduces the disk-integrated, time-averaged line profiles of the Institut Für Astrophysik, Göttingen (IAG) Solar Atlas \citep{2016A&A...587A..65R}, but it has not, prior to our work, been tested against highly precise RV observations.

While granulation-induced CLV of line profiles can be directly observed on the Sun due to its spatially resolved surface, such information can only be indirectly probed for other stars using planetary transits \citep{2017A&A...605A..90D, 2017A&A...605A..91D, 2018A&A...616A.144D}. Using archival data of HD 209458 and HD 189733 from the UVES \citep{2000SPIE.4008..534D} and HARPS \citep{2000SPIE.4008..582P} spectrometers, \cite{2017A&A...605A..91D, 2018A&A...616A.144D} extract center-to-limb observations of Fe I lines with remarkable accuracy to predictions from hydrodynamic 3D models. Yet, the Sun, with its spatially resolved surface and precisely known dynamics, provides an ideal opportunity to test the accuracy of RV models that account for the granulation signatures. During a solar eclipse, solar RVs show the Rossiter-McLaughlin \citep[RM,][]{1924ApJ....60...15R, 1924ApJ....60...22M} effect, which is characterized by blue/red-shifted deviations in the RV curve that scale with the fraction of surface occulted, the star's rotational velocity, and the spin-orbit alignment between the stellar spin axis and the planet's orbital plane \citep{2007ApJ...655..550G}. Estimates of spin-orbit alignment angles inform planet migration theories, as the degree of misalignment serves as a proxy for the degree of violent interactions during migration. However, ignoring granulation-driven convective blueshift in RM observations leads to RV deviations, which impact estimates of spin-orbit alignment angles \citep{2011ApJ...733...30S, 2016A&A...588A.127C, 2016ApJ...819...67C}. A solar eclipse provides an unique opportunity to investigate the influence of the granulation signatures on RVs, as the large occultation from the Moon results in a large RM effect and drastically altered spectral line shapes.   

The first study of this manner was performed by \cite{2016A&A...595A..26R} for the 2015 March 20 solar eclipse, which was observed with the IAG Fourier Transform Spectrograph \citep{2016PASP..128i5002L}. With these observations, \cite{2016A&A...595A..26R} tested various RV models with distinct treatments of convective blueshift to highlight the influence of convective blueshift on the RM curve. They found that inclusion of convective blueshift in their RV models reduced the peak velocity residual by roughly 50\% regardless of how the RVs are computed. However, even with convective blueshift accounted for, there still remained a significant data-model discrepancy with peak residuals up to $\sim$25 m s$^{-1}$. They suspected their models were limited by lack of understanding line profile CLV, for which they adopted a sigmoid function with arbitrarily prescribed coefficients. Here we follow up on the \cite{2016A&A...595A..26R} study by modeling the 2023 October 14 solar eclipse using \texttt{GRASS} to determine whether the empirically based treatment of granulation therein is able to better replicate solar eclipse RVs. 

Spectral observations of the Sun as a star were obtained during the 2023 October 14 solar eclipse using the NEID solar feed \citep{2022AJ....163..184L} and spectrograph \citep{2016SPIE.9908E..7HS,Halverson2016} at Kitt Peak National Observatory. We model the solar RVs using parameterizations for limb darkening, differential extinction, differential rotation, and the granulation information from \texttt{GRASS}. The data used are described in \S\ref{sec:Data}. The modifications to \texttt{GRASS} and the RV models used are discussed in \S\ref{sec:models}, followed by a discussion of our results in \S\ref{sec:Results}.

\begin{figure*}[!htb]
    \includegraphics[width=1.0\textwidth, trim=0cm 13cm 0cm 13cm, clip, height=0.25\textwidth]{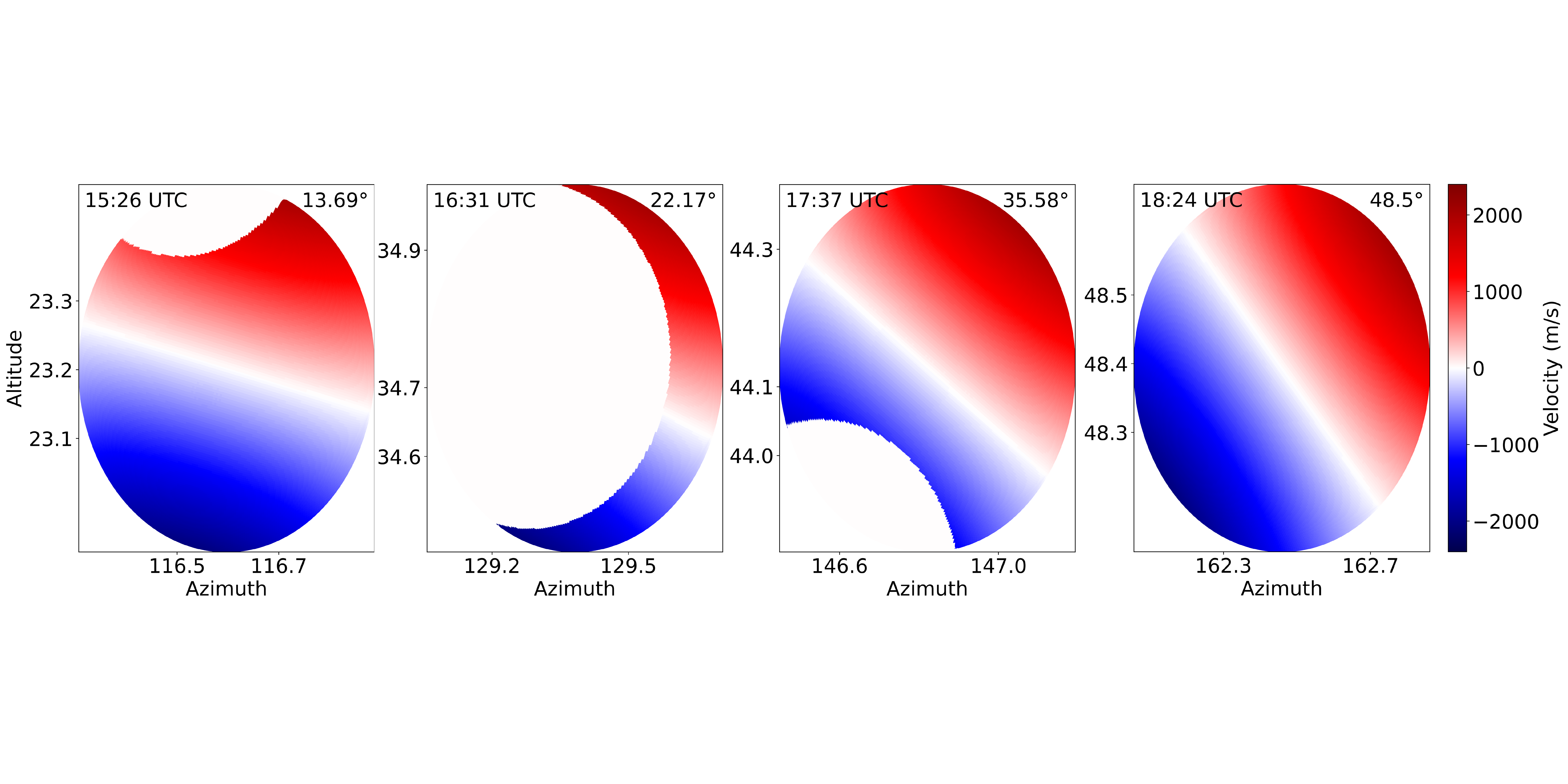}
    \caption{On-sky configuration of solar eclipse as viewed from NEID via simulated Dopplergrams, with colors depicting radial velocities from solar differential rotation only. UTC time is given in the upper left panel corner and the angle between horizon and solar rotation axis is in the upper right panel corner.}
    \label{fig:eclipse_plots}
\end{figure*}

\section{Eclipse Observations}
\label{sec:Data}
Solar eclipse RVs are sensitive to granulation signatures and so allow us to test the accuracy of granulation-informed RV models. However, such RV observations require high spectral resolution in order to best capture the granulation signatures. We designed an experiment to obtain such data with the NEID spectrometer, and describe the obtained observations below. 

\subsection{Instrument}
The 2023 October 14 solar eclipse was observed with the NEID spectrograph \citep{2016SPIE.9908E..7HS}, which is attached to the WIYN 3.5~m telescope\footnote{The WIYN Observatory is a joint facility of the NSF National Optical-Infrared Astronomy Research Laboratory, Indiana University, the University of Wisconsin-Madison, Pennsylvania State University, and Princeton University.} at Kitt Peak National Observatory. NEID is a state-of-the-art high-resolution spectrometer ($R \sim 115{,}000$, 380 -- 930 nm) that obtains spectroscopic data of nearby stars. NEID is highly stabilized for an instrumental RV precision of better than 27 cm s$^{-1}$ \citep{Halverson2016} and is wavelength-calibrated using a laser-frequency comb and Fabry-Pérot etalon. This analysis specifically makes use of the NEID solar feed \citep{2022AJ....163..184L} for Sun-as-a-star observations. The NEID solar feed assembly includes a solar tracker to maintain pointing precision and a pyrheliometer as a cloud sensor. More details of the NEID solar feed can be found in \cite{2022AJ....163..184L}. Sun-as-a-star NEID observations over the course of 3.5 years when modeled with \texttt{SCALPELS} \citep{2021MNRAS.505.1699C} reveal an on-sky RV precision of 37 cm s$^{-1}$ --- suggesting that the capability of detecting Earth-mass exoplanets orbiting solar-mass stars may be coming into reach \citep{2024arXiv240813318F}.

\subsection{Observations}
At its maximum, the 2023 October 14 solar eclipse covered a total of 76\% of the solar disk when viewed from Kitt Peak. The eclipse began at 15:11 UTC (8:11 AM local time) at an airmass of 2.86 and ended at 18:03 UTC (11:03 AM local time) at an airmass of 1.38. The maximum occurred at 16:31 UTC (9:31 AM local time). The eclipse configurations are shown in Figure \ref{fig:eclipse_plots}. NEID solar observing automatically triggers at 9:30 AM local time, which is after the start time of the eclipse. Therefore, observations were manually triggered starting at 8:04 AM local time immediately after a shortened (high-resolution mode only) calibration sequence following the on-sky observing from the previous night. Due to clouds at the beginning and end of the observations, our analysis data begin at 15:26 UTC (8:26 AM local time) and ends at 18:24 UTC (11:24 AM local time) for a total of 130 observations. The exposure time of each observation followed the standard NEID solar observation sequence of 55 seconds exposure and a read out time of 27 seconds for a total observing cadence of 82 seconds. We note that daily at 10 AM local time a liquid nitrogen fill occurs that does induce a change in the intrinsic velocity drift of the NEID spectrometer \citep{2022AJ....163..184L}. However, this drift is tracked and removed by a simultaneous calibration source observed with the solar data. The measured signal-to-noise ratio (SNR) per resolving element of the observed spectra for this work's case study line, 5434.5 $\text{\AA}$ (air wavelength, vacuum wavelength of 5436.0 \AA), is $\sim$378 outside of the eclipse and drops to $\sim$161 during the time of maximum eclipse. During normal operations an automated routine periodically ensures that the output image of the fiber carrying the solar light is well centered on the NEID input fiber at the telescope port. However, during this experiment we opted not to engage this routine to prevent issues due to the rapidly changing intensity of the Sun from the eclipse.

\subsection{Measuring Radial Velocities}
\label{RVs}
The NEID data reduction pipeline\footnote{\url{https://neid.ipac.caltech.edu/docs/NEID-DRP/index.html}} (DRP) RVs are calculated by computing a cross-correlation function (CCF) between each observed spectrum and publicly available ESPRESSO \citep{2021A&A...645A..96P} masks. 
Since the NEID DRP RVs are essentially an average RV computed from a large array of lines simultaneously, they do not allow for investigation of line-by-line variability. Therefore, we separately compute line-by-line RVs from the NEID observed spectra for a selected set of lines between 525--630~nm. We use \texttt{RvSpectML} \citep{eric_ford_2021_4593963}, which provides multiple algorithms for estimating RVs from a CCF, to extract a specific spectral line from the NEID spectra and compute line-by-line RVs following the CCF algorithm from \cite{2002A&A...388..632P}. The CCF is computed by shifting a template mask along the spectra at varying velocities. Our template masks are top-hat functions centered at the rest-frame wavelength of each individual spectral line with the height determined by the spectral line depth. Following \cite{2002A&A...388..632P}, we determine the RV from the CCF by fitting a Gaussian around the CCF minimum (defined using 75\% of the CCF full width at half maximum) using the Levenberg-Marquardt algorithm. The mean of the Gaussian fit represents the apparent Doppler velocity of the spectral line and therefore is our line-by-line RV. The same procedure is used to compute the simulated RVs described in Section \S\ref{subsec:rv_models}. 

\section{Eclipse Simulations}
\label{sec:models}
The spectral signatures of granulation are modeled using a modified version of \texttt{GRASS} v2.0. \texttt{GRASS} creates synthetic spectra with granulation-driven variability from spatially and temporally resolved observations of 22 solar absorption lines \citep{2022AJ....163...11P,2024AJ....168...46P}. Spectroscopic parameters for these spectral lines can be found in Table 1 in \cite{2024AJ....168...46P}. Each input spectral line used by \texttt{GRASS} v2.0 was observed by LARS ($R \sim 700{,}000$, \citealt{2017A&A...607A..12L}) at up to eleven positions along the North, South, East and West axes from disk center with a field of view up to 40x30 arcseconds and a pointing accuracy of $\sim$1 arcsecond \citep{2018A&A...611A...4L, 2019A&A...622A..34S, 2019A&A...624A..57L}. Such spatially resolved observations are critical when attempting to capture CLV. The observations were done in $\sim$20 min sequences with each exposure having a cadence of 1.5 s then binned into ten exposures (15 s). By using these data, \texttt{GRASS} captures temporal variability in line shape and convective blueshift. Furthermore, the statistical behavior of granulation used in \texttt{GRASS} has been shown to closely match state-of-the-art hydrodynamic simulations \citep{2025MNRAS.539.2248F}. We refer the reader to \cite{2022AJ....163...11P} and \cite{2024AJ....168...46P} for specific details on the spatial interpolation and integration procedure used by \texttt{GRASS}. 

\subsection{\texttt{GRASS} Repurposed for Solar Observations}
\label{sec:Modifications}
\texttt{GRASS} was intentionally created for stars and makes extensive use of large-distance approximations that assume the observing geometry does not drastically change with time. Therefore, as written, \texttt{GRASS} cannot be used to model the resolved Sun from the perspective of the Earth orbiting it. To use \texttt{GRASS} to interpret solar observations, we modify the geometry (\S \ref{sec:geometry}), explore various new intensity models (\S\ref{sec:LD}, \S\ref{sec:ext}) and incorporate sunspots (\S\ref{sec:spots}). These modifications are available on the \texttt{GRASS} GitHub\footnote{\url{https://github.com/palumbom/GRASS}} under the Julia submodule \texttt{GRASS.Eclipse}, referred to as  \texttt{GRASS-E} herein. 

\subsubsection{Geometry Modifications}
\label{sec:geometry}
The geometry used in \texttt{GRASS-E} generates synthetic granulation-driven spectra of the eclipsed Sun. The state vectors of the Earth, Moon and Sun during the eclipse are acquired using NASA’s Navigation and Ancillary Information Facility SPICE toolkit \citep{ACTON199665}, which provides highly accurate ephemerides of Solar System objects with an accuracy of a few cm s$^{-1}$. Such state vectors are corrected for light travel time within the SPICE toolkit by iteratively solving the light time equation until a solution converges. Both the Moon and Sun are assumed to be perfect spheres; whereas, Earth’s oblateness, defined as the fractional difference between the equatorial and polar radii, is accounted for in calculating the position of the observatory. We maintain the same solar surface gridding algorithm described in Appendix A.2 in \cite{2024AJ....168...46P}. For each cell on the solar grid, the differential rotation velocity is determined using the following functional form: 
\begin{equation} \label{eq:dr}
      \omega(l) = A - B \sin^2(l) - C \sin^4(l)
\end{equation}
where the angular rotational velocity $\omega$ is a function of solar latitude $l$ \citep{1970SoPh...12...23H}. 
We use the spectroscopic coefficients from \cite{1990ApJ...351..309S} ($A = 14.113^{\circ}$ day$^{-1}$, $B = -1.698^{\circ}$ day$^{-1}$, $C = -2.346^{\circ}$ day$^{-1}$).
The differential rotation velocity is projected onto the line of sight from each individual solar cell to the observer, which is specified by user-given latitude, longitude and altitude\footnote{NEID solar feed latitude: 31.9583\textdegree N; longitude: 111.5967\textdegree W; altitude: 2096 m}. In order for our models to best mimic the involved dynamics under which the observations were obtained, we include the velocity from the Earth's rotation and orbital motion towards each individual solar cell. Hence, the final velocity of each solar cell is the sum of the projected differential rotation velocity and the projected rotation and orbital velocity of the Earth towards the solar cell. By including the barycentric motion of the observer on Earth in our models, we can apply a consistent barycentric correction to both the NEID-observed and simulated RVs using \texttt{barycorrpy} \citep{2018RNAAS...2....4K}. \par 

The key solar-related modification in \texttt{GRASS-E} is that for each epoch we must update the SPICE query of the Earth, Moon, and Sun state vectors, which are used to determine (a) whether a solar grid cell is eclipsed or not, and (b) the time-evolving projected cell velocity and area grid. Based on this new geometry, \texttt{GRASS-E} generates time-series absorption line profiles that are shifted by each solar cell's projected velocity, then integrated (weighted by projected area and intensity) to produce synthetic flux spectra. \par  

\subsubsection{Limb Darkening Modifications}
\label{sec:LD}
In addition to geometry modifications, we also modify the intensity model in \texttt{GRASS-E}. The default limb darkening model used by \texttt{GRASS} is the wavelength-independent, quadratic limb darkening law (Equation~\ref{eq:1}) from \cite{1950HarCi.454....1K} with fixed coefficients of u$_1 = 0.4$ and u$_2 = 0.26$ \citep{2008MNRAS.386.1644S}. 
\begin{equation} \label{eq:1}
    \frac{I(\lambda,\mu)}{I(\lambda,\mu=1)} = 1 - u_1(1-\mu)- u_2(1-\mu)^2
\end{equation}
However, commonly used limb darkening laws, such as the quadratic law, fail to reproduce true center-to-limb intensity variations \citep{2022A&A...662A..38N}. \cite{2024NatAs...8..929K} show that magnetic fields have a direct impact on limb darkening and can help explain the discrepancies between stellar atmosphere models and observations. \cite{2024NatAs...8..929K} use MURaM \citep{2005A&A...429..335V}, a three-dimensional radiative MHD code, to simulate limb darkening values at $\mu = 0.1$ -- $1.0$ (with a $\Delta \mu$ of 0.1) for the following magnetic field strengths: hydrodynamic (i.e. no field), small-scale turbulent dynamo (SSD), 100 G, 200 G, and 300 G. The SSD generates a baseline level of magnetic flux (70 G) across the entire stellar surface as a result of convective motion distorting magnetic field lines \citep{2020ApJ...894..140R}. 

\begin{figure}[!tb]
    \epsscale{1.25}
    \plotone{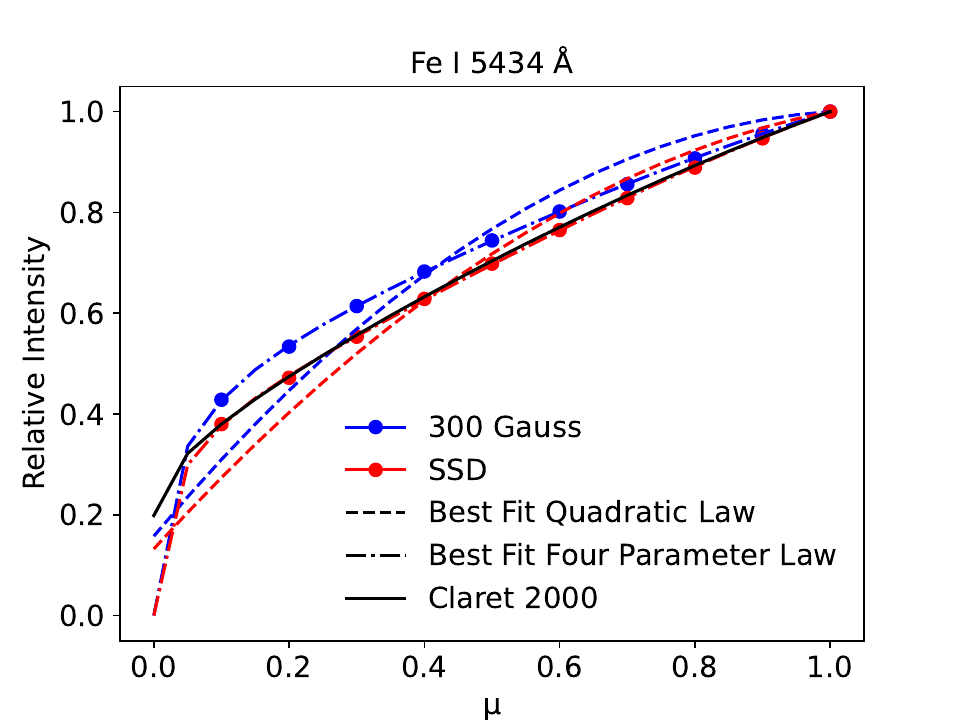}
    \caption{The simulated limb darkening at SSD (red) and 300 G (blue) from \cite{2024NatAs...8..929K} used for Fe I 5434 $\text{\AA}$ as a function of $\mu$. The best fit quadratic (dashed) and four-parameter (dot-dashed) laws are shown in the corresponding magnetic field strength color. The solid black line shows the V-band four-parameter law as determined by \cite{2000A&A...363.1081C} for the Sun. For both featured magnetic field strength, the fitted four-parameter law (dot-dashed) matches the simulated limb darkening best.}
    \label{fig:LD}
\end{figure}

The four-parameter limb darkening law is considered the most accurate \citep{2023A&A...671A.130P} and is as follows:
\begin{equation} \label{eq:2}
    \frac{I(\lambda,\mu)}{I(\lambda,\mu=1)} = 1 - \sum_{i=1}^{4}u_i(1-\mu^{i/2}).
\end{equation}
We fit the quadratic (Equation~\ref{eq:1}) and four-parameter (Equation~\ref{eq:2}) limb darkening laws to the SSD and 300 G MURaM simulated limb darkening. 
Figure \ref{fig:LD} shows the simulated limb darkening values as a function of $\mu$ at SSD and 300 G taken from \cite{2024NatAs...8..929K} for 5428.5 $\text{\AA}$, which is the closest to our case study line of Fe I 5434 $\text{\AA}$ (air wavelength). The best fit quadratic and four-parameter laws are shown in the dashed and dot-dashed lines, respectively, in the corresponding magnetic field strength color. In addition, we include in the solid black line the V-band four-parameter law as determined by \cite{2000A&A...363.1081C}. The \cite{2000A&A...363.1081C} limb darkening prescription is based on plane-parallel stellar atmosphere models, which fail to capture the impact of magnetic fields on limb darkening. As shown in Figure \ref{fig:LD}, the \cite{2000A&A...363.1081C} limb darkening prescription aligns with the SSD simulated limb darkening, which differs from the 300 G simulated limb darkening. This is consistent with the \cite{2024NatAs...8..929K} finding that stellar atmosphere models are only able to reproduce limb darkening when little to no magnetic fields are physically involved.  \cite{2024NatAs...8..929K} found that for the quiet Sun the SSD limb darkening is the most appropriate. 
Since \texttt{GRASS-E} is informed by quiet-Sun LARS observations, we decide to use the simulated SSD limb darkening values to determine our limb darkening model. However, we do caution that the solar eclipse occurred when the Sun was near a maximum activity phase, so using SSD limb darkening and quiet Sun center-to-limb observations to inform CLV may not capture all magnetic effects. 

Given its flexibility near-limb, we use the four-parameter law fit of the simulated SSD limb darkening over the quadratic law. We find that although the \cite{2000A&A...363.1081C} V-band limb darkening prescription aligns well with the Fe I 5434 $\text{\AA}$ simulated SSD limb darkening values (Figure \ref{fig:LD}), towards the redward end of the V-band the prescription starts to fail. For example, the reddest template line from \texttt{GRASS-E}, Fe I 6302 $\text{\AA}$, has a growing deviation towards smaller $\mu$ between the \cite{2000A&A...363.1081C} V-band limb darkening prescription and the \cite{2024NatAs...8..929K} simulated SSD limb darkening. This suggests that line-by-line limb darkening is preferred over the V-band limb darkening prescription. Hence, our intensity models use the four-parameter limb darkening law (Equation~\ref{eq:2}) with the line-by-line fitted coefficients from the simulated SSD limb darkening.   

\subsubsection{Modeling Extinction}
\label{sec:ext}
\begin{figure}[!tb]
    \epsscale{1.25}
    \plotone{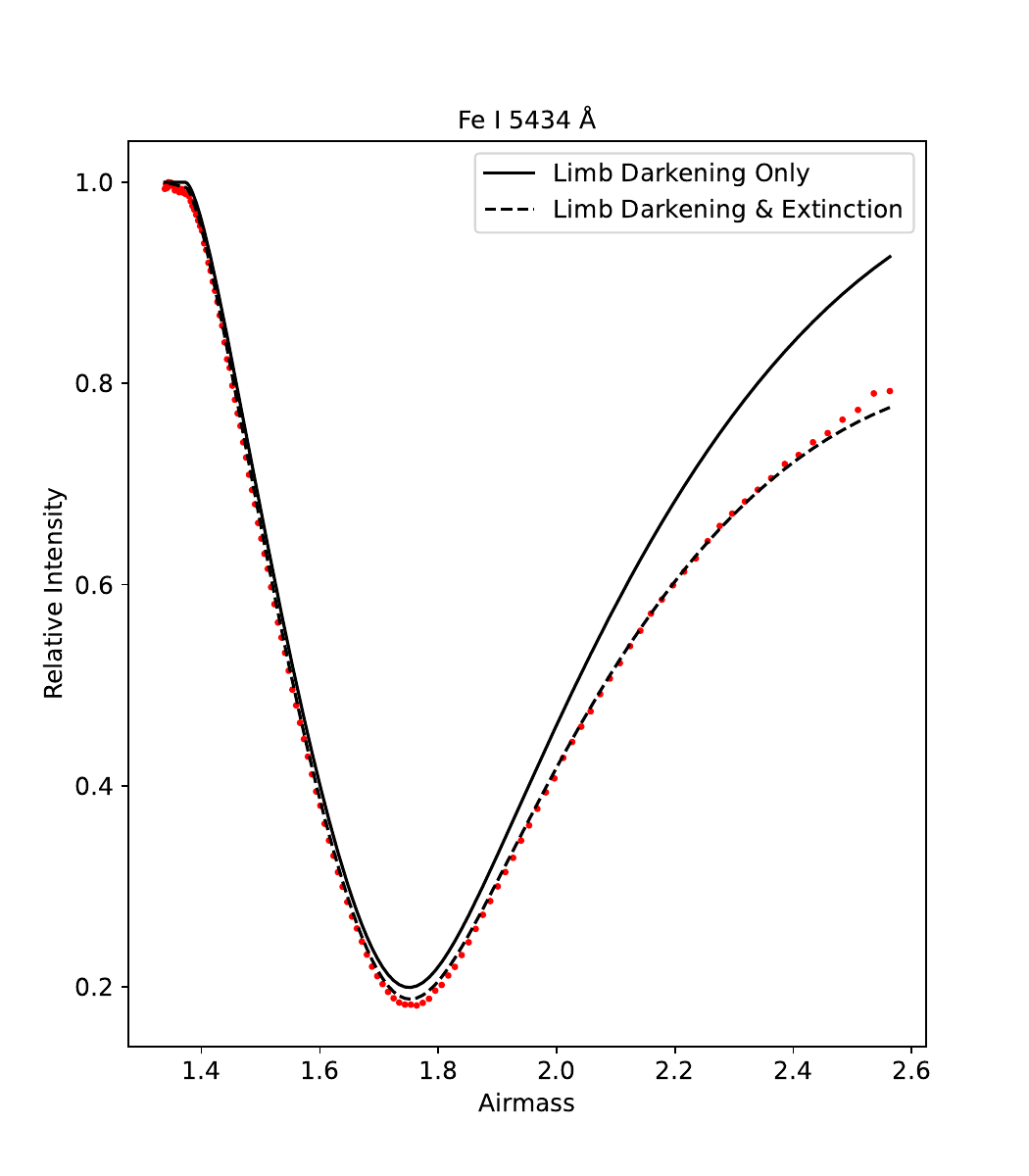}
    \caption{The observed NEID relative intensity (dotted red line) and simulated relative intensity with only the SSD four-parameter limb darkening (solid line) and including extinction (dashed line) for Fe I 5434 $\text{\AA}$ as a function of airmass during the October 2023 solar eclipse. Unsurprisingly, including both limb darkening and extinction in our models produces a better fit to the observed data, especially at higher airmass values. }
    \label{fig:extinction}
\end{figure}

In addition to updating the limb darkening model used in \texttt{GRASS-E}, we also include differential extinction in our intensity model. Since the eclipse occurred early in the morning while the Sun was at a high airmass, we expect extinction to be strong. In addition, due to the Sun's large angular size (0.5 degrees),  extinction will differ significantly across the solar disk. Under cloud-free conditions, extinction follows an exponential law as a function of airmass \citep{2019MNRAS.487.1082C}. The corresponding extinction coefficients are seeing-dependent, and so are generally unknown. 
To determine the wavelength-dependent extinction coefficients ($\alpha(\lambda)$), we perform a non-linear least-squares fit (via the Levenberg-Marquardt algorithm) of the following: 

\begin{equation} \label{eq:3}
    I_\text{NEID}(\lambda,t) = I_\text{LD}(\lambda,t) \times e^{-\alpha(\lambda) \times \sec(z)}
\end{equation}

\noindent where $I_\text{NEID}$ is the observed NEID intensity, $I_\text{LD}$ is the four-parameter limb darkening law (Equation~\ref{eq:2}) as a function of wavelength and time, and $z$ is the zenith angle. Figure \ref{fig:extinction} highlights the impact of including extinction in our intensity model: including the best-fit extinction reduced the intensity residuals by nearly 10x. We confirm that our estimated wavelength-dependent extinction coefficients are reasonable and similar in value to the computed extinction coefficients by \cite{2024arXiv240813318F} for solar NEID observations. We also allow the extinction coefficients to slightly vary with time to reflect extinction changes during the eclipse and find no significant improvements in our models. Hence, we proceed with constant-in-time wavelength-dependent extinction coefficients fitted from the extinction law.

\subsubsection{Modeling Sunspots}
\label{sec:spots}
\begin{figure}[!htb]
    \epsscale{1.05}
    \plotone{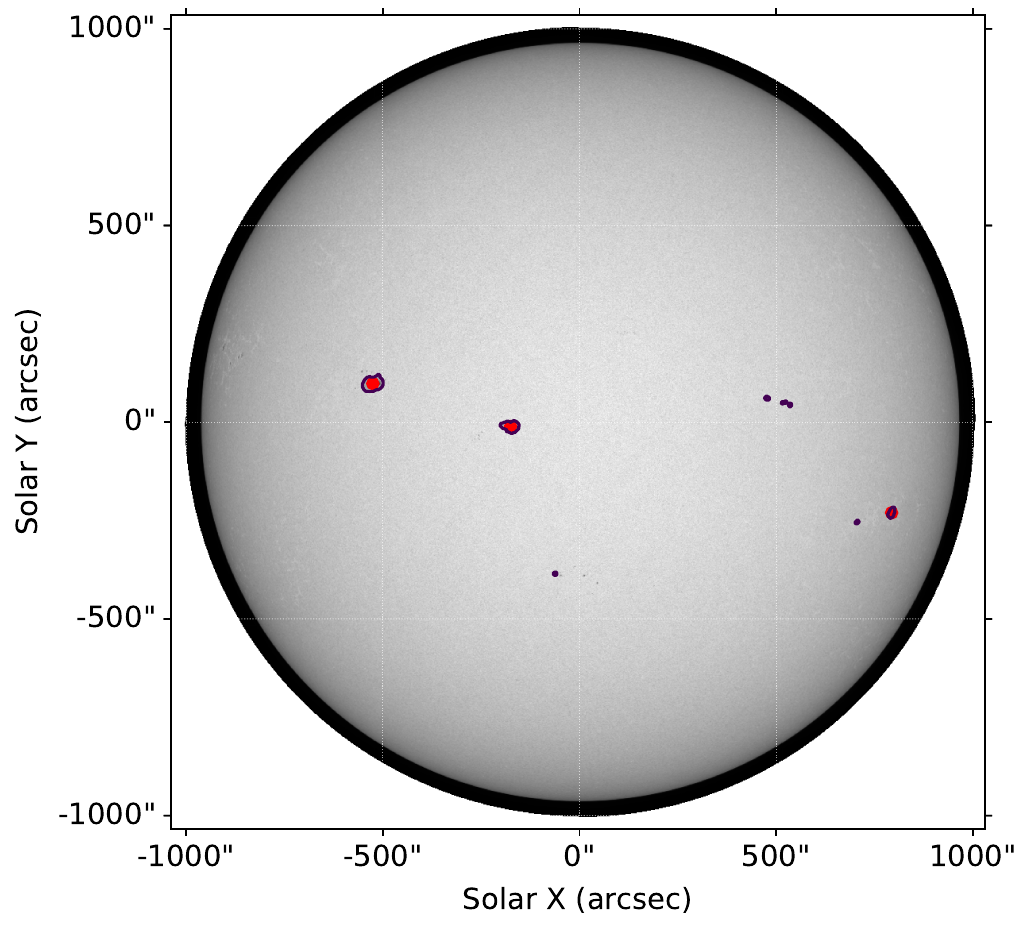}
    \caption{HMI continuum images taken on 2023 October 14 at 16:35 UTC. All sunspots present are depicted. The largest three sunspots are highlighted in red and included in our RV models.}
    \label{fig:sunspot}
\end{figure}

Magnetically active regions, such as spots, degrade RV precision in two distinct ways: (1) the flux effect and (2) the suppression of convective blueshift effect. The flux effect arises from the temperature contrast between an active region and the surrounding quiet photosphere \citep{2014ApJ...796..132D}. These active regions emit different amounts of light than the surrounding quiet photosphere, breaking the symmetry between the rotationally red- and blue-shifted hemispheres and introduce asymmetries in disk-integrated spectral line profiles and quasi-sinusoidal RV signals. The suppression of convective blueshift effect results from magnetic fields within an active region suppressing convective motion \citep{1985A&A...143..116C}. In the Sun, this suppression can reach $\sim$300 m s$^{-1}$ in spots, making magnetically active regions appear redshifted relative to the average photosphere \citep{2010A&A...512A..39M}. Images taken by the Helioseismic and Magnetic Imager (HMI, \citealt{2012SoPh..275..229S}) onboard the Solar Dynamics Observatory \citep[SDO,][]{2012SoPh..275....3P} reveal the presence of sunspots during the 2023 October 14 solar eclipse. Figure \ref{fig:sunspot} shows the downloaded HMI continuum image (exposure time of 720 s) at 16:35 UTC, near maximum eclipse. All sunspots are shown in Figure \ref{fig:sunspot} with the largest three sunspots highlighted in red. The HMI image is used to extract the coordinates, effective radius, and mean intensity contrast for each of the largest three sunspots. Given this information, we are able to include both the flux and convective blueshift effects in our RV model. We assume the spots were circular with no intensity gradient and not changing in size or position. Based on the spots' coordinates and angular sizes, cells in our solar grid that are found inside a spot have their intensity reduced by the determined intensity contrast and their convective blueshift suppressed to zero. We find that accounting for the flux and convective blueshift effects only improves our RV residuals RMS by $<1.0$ m s$^{-1}$. We acknowledge that other sources of magnetically active regions, such as plages, can impact RV precision. However, given the minuscule improvement in RV RMS when including spots, we do not explore other sources of magnetically active regions. 

\subsection{Radial Velocity Models} 
\label{subsec:rv_models}
We simulate the 2023 October 14 solar eclipse RVs at the mid-exposure time of the NEID observations using three distinct models. We find that during the 55 second NEID exposure, solar flux from the NEID exposure meter varies linearly throughout the eclipse. In addition, we find that simulating the RVs at fine samples (down to one second) within the 55 second exposures and then computing a weighted average RV for an individual 55 second exposure does not significantly improve our models. Each model follows the same geometry and intensity computations described within \S \ref{sec:Modifications}. 
\begin{itemize}
\item{Model I computes RVs as the weighted average projected velocity by weighting each solar cell's projected velocity with its corresponding intensity and projected area. This model does not account for any granulation signatures.}  
\item{Model II computes RVs using the CCF-algorithm described in Section \ref{RVs} on simulated \texttt{GRASS-E} spectra that exclude all time-varying granulation signatures (convective blueshift jitter, line asymmetry).}
\item{Model III also computes CCF RVs but from \texttt{GRASS-E} simulated granulation-driven spectra.}
\end{itemize}
Model III is the only model that runs the original \texttt{GRASS} granulation-driven spectra simulator. While model II is run through \texttt{GRASS-E} for consistency in surface gridding, spectra simulation, and CCF RV computation, the generated spectra from model II do not account for any granulation signatures. In other words, model II grids the Sun as model III does, but rather than assigning a line with granulation signatures to each cell, model II instead assigns a symmetric non-blueshifted line that is static with time. From there, model II performs a spatial integration to get a disk-integrated spectrum (no granulation signatures) from where a CCF RV is computed. Since the only difference between models II and III is the inclusion of the granulation signatures, we can use these models to quantify the impact of granulation and validate \texttt{GRASS-E}'s granulation modeling. We note that the observations informing \texttt{GRASS-E}'s granulation model are not contemporaneous with the solar eclipse. Therefore, while model III captures the statistical behavior of granulation, the asynchronous nature of the data acts as uncorrelated noise at short timescales. However, the intrinsic granulation-induced RV variability is on the order of $\sim$45--75 cm s${-1}$ \citep{2024AJ....168...46P}, which is small compared to our photon noise limit of $\sim$2--3 m s$^{-1}$ for individual lines. Hence, the temporal mismatch contributes minimally to the overall RMS and does not drive the larger structures seen in the residuals.

In order to directly compare the RVs from models II and III to the NEID line-by-line RVs, the \texttt{GRASS-E} simulated spectra must be degraded from the resolving power set by the LARS input data ($R \sim 700{,}000$) down to NEID's resolving power of $R \sim 115{,}000$. We test three line spread functions (LSF): a Gaussian, a symmetric NEID LSF\footnote{\url{https://neid.ipac.caltech.edu/docs/NEID-DRP/algorithms.html\#variable-line-spread-function}}, and an asymmetric NEID LSF (Joshi in prep.). The NEID LSF is a non-Gaussian kernel and is modeled as a convolution of: the pixel response function, a 1D integrated circular optical fiber image, and a Gaussian with a standard deviation of 0.6 pixels.  However, significant residuals are observed when comparing this symmetric LSF model with the laser frequency comb (LFC) spectrum. Joshi in prep.\ account for both the symmetric and asymmetric deviations from the observed LFC spectrum by perturbing the model LSF with orthogonal polynomials up to degree 8 in order to model the true NEID LSF. We find that the two NEID LSF models perform comparably and tend to outperform the Gaussian LSF. Therefore, we use the asymmetric NEID LSF to degrade the \texttt{GRASS-E} spectra to NEID's resolving power. 

All three models are computed separately for each of the 22 \texttt{GRASS-E} template lines in order to incorporate wavelength-dependent limb darkening, extinction, and granulation signatures. Line-by-line models allow us to evaluate the variability of each line (see Section \S \ref{sec:line_var}). We compare the observed and simulated line profiles of the 22 \texttt{GRASS-E} template lines and confirm that, within 2\%, \texttt{GRASS-E} successfully reproduces the line profiles from the NEID spectra. 

\section{Results}
\label{sec:Results}
The RM curve of the 2023 October 14 solar eclipse was simulated using the three models described in Section \S\ref{subsec:rv_models}. We here use the Fe I 5434 $\text{\AA}$ spectral line as an initial case study. 

\begin{figure}
    \centering
    \includegraphics[scale=0.5]{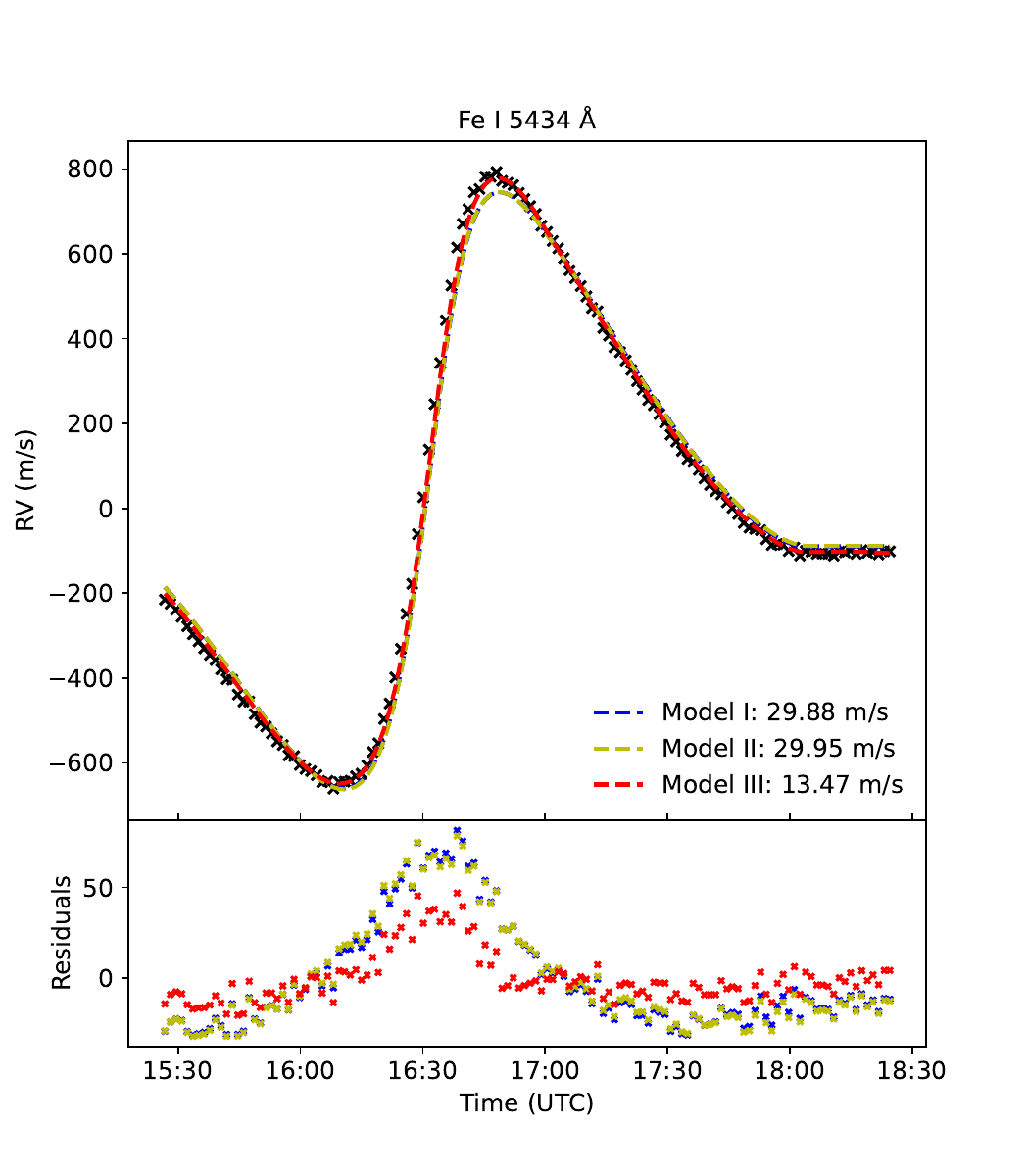}
    \caption{Top panel: black crosses show the NEID line-by-line RVs for the Fe I 5434 $\text{\AA}$ line. Dashed blue line shows model I, average projected velocities with no convective blueshift. Dashed yellow line shows model II, CCF RVs from simulated spectra with no granulation signatures. Dashed red line shows model III, \texttt{GRASS-E} CCF RVs. All RVs are mean-subtracted. Residual RMS after binning into 4 exposures (5.4 min) is found in legend. Bottom panel: residuals between the observations and each model in the corresponding color. Model III, which includes granulation signatures, performs better than the models that do not include granulation signatures.}
    \label{fig:default_models}
\end{figure}

Figure \ref{fig:default_models} shows the NEID line-by-line RVs for Fe I 5434 $\text{\AA}$ along with the simulated RVs from each model. Models I and II show similar RM curves with an offset towards the blue (negative velocities) during the eclipse. The simulated RM curves of models I and II are significantly similar based on their residuals following the same shape (peaking at $\sim$75 m s$^{-1}$) and their binned RMS both at $\sim$30 m s$^{-1}$. To average over the $\sim$5.4 min solar p-mode period, we bin RVs into sets of 4 exposures (328 s, 5.4 min). Neither model I nor II accounts for any granulation signatures. The key difference between models I and II is how the RVs are computed: model I computes an average weighted projected velocity across the solar disk, while model II computes a CCF RV from simulated spectra. Given the similarities between the RM curves of models I and II, how the RVs are computed does not seem to significantly influence the shape of the simulated RM curve. Model III, which computes a CCF RV from \texttt{GRASS-E} simulated granulation-driven spectra, performs better than models I and II by a factor of more than 2 in RV RMS. Figure \ref{fig:default_models} highlights that model III does experience a similar offset towards the blue like the other two models. The key difference between models II and III (both CCF RVs from simulated disk-integrated spectra) is that model III accounts for convective blueshift and line asymmetries, whereas model II does not account for any granulation signatures. As a result, model III's better performance can be attributed to its inclusion of granulation signatures. The results from these models emphasize the known need to account for granulation signatures in solar eclipse RV models \citep{2016A&A...595A..26R}. 

Following \cite{2001A&A...374..733B}, we estimate the photon noise for Fe I 5434 $\text{\AA}$ to be: 2.75 m s$^{-1}$ for all considered observations, 3.01 m s$^{-1}$ during the eclipse, and 2.05 m s$^{-1}$ outside of the eclipse. Outside of the eclipse is defined as 18:05 UTC up to 18:24 UTC, where clouds start to affect the RVs. We obtained a total of 15 out-of-eclipse observations, each with a cadence of 82 s. 
Our full data binned RMS (depicted in Figure \ref{fig:default_models}) do not come close to the photon noise limit, suggesting that there may be missing or incorrect astrophysics in our RV models. We explore some potential sources of errors in the following sections. 

\subsection{Pressure Modes}
\label{sec:pmodes}
Pressure modes, or p-modes, are oscillations on a stellar surface that arise due to traveling sound waves that are generated by pressure fluctuations from convection in the interior of stars. Although p-modes introduce $\sim$m s$^{-1}$ RV variability, they can be averaged down to sub-10 cm s$^{-1}$ level by binning RVs according to the principal p-mode period \citep{Chaplin2019}, which is $\sim$5.4 minutes for the sun. Hence, we bin the entire eclipse RVs into sets of 4 exposures ($\sim$5.4 min). We find that binning reduces the RMS very slightly, suggesting p-modes are involved. However, the improvement is very small compared to the residuals. We test how RMS changes as a function of different bin sizes and find that additional binning does not contribute to a significant reduction in RMS. Hence, p-modes alone cannot explain the data-model discrepancy during the eclipse. We do acknowledge that during the eclipse partial blocking of the solar disk occurs and results in the symmetry breaking of the p-mode averaging; therefore, it is unclear whether averaging over a 5.4 min period continues to successfully remove p-mode variability almost entirely, as expected in precision RV surveys \citep{Chaplin2019, 2021AJ....161..130G}. For consistency with how a normal star would be observed in such surveys, we present all our RV RMS with this 5.4 minute binning, though our plots show the individual 55 second RV observations. 

\subsection{Limitations of \texttt{GRASS-E} }
The \texttt{GRASS-E} treatment of granulation signatures is informed by the center-to-limb observations of 22 solar absorption lines taken by LARS \citep{2018A&A...611A...4L, 2019A&A...622A..34S, 2019A&A...624A..57L}. Each absorption line was observed at up to eleven heliocentric positions along four radial axes (North, South, East and West) with sampling of $\mu$ = [0.2, 0.3, 0.4, 0.5, 0.6, 0.7, 0.8, 0.85, 0.9, 0.95, 1.0], but for some lines only up to $\mu$ of 0.3.
\begin{figure}
    \centering
    \includegraphics[scale=0.55]{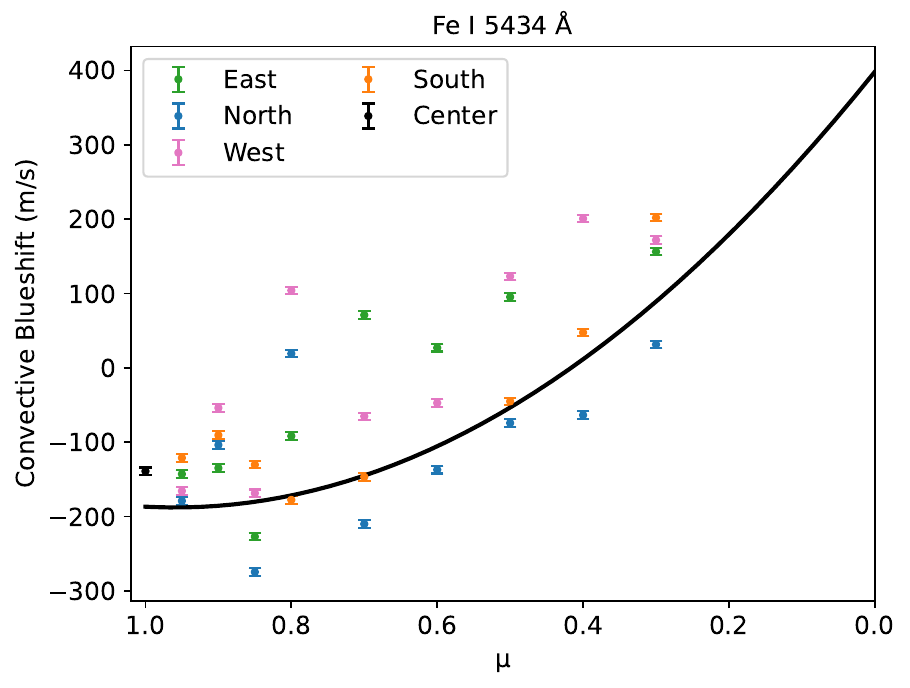}
    \caption{\texttt{GRASS-E} convective blueshift measurements as a function of $\mu$ for Fe I 5434 $\text{\AA}$ taken by LARS \citep{2019A&A...624A..57L}. Scattered dots depict the convective blueshift at a given $\mu$ on a specific axes: North-blue, East-green, South-orange, and West-pink. Error bars indicate the mean error of 5 m s$^{-1}$. LARS observations used by \texttt{GRASS-E} extend only to a minimum $\mu$ of 0.3 for Fe I 5434 $\text{\AA}$. As shown, the convective blueshift measurements used by GRASS-E are non-continuous and do not cover near-limb. Black line shows the mean-subtracted optimized second-order polynomial ($a \mu^2 + b \mu$) for the convective blueshift of Fe I 5434 $\text{\AA}$.}
    \label{fig:cbs}
\end{figure}
For each LARS observation, \texttt{GRASS-E} computes and stores the line convective blueshift, along with the line bisector and width, as a function of line depth. 
This information is then used to generate synthetic disk-resolved line profiles. Although the LARS observations used by \texttt{GRASS-E} have a pointing accuracy of $\sim$1 arcsecond and a high resolving power of $\sim$700,000, the field of view of the observations covers up to 40x30 arcseconds in order to average over the approximate size of a supergranule. Such large field of view introduces uncertainties of the observed heliocentric position ($\mu$), which worsen towards the limb. As a result, the measured CLV of bisector asymmetry and convective blueshift used by \texttt{GRASS-E} are for a superposition of $\mu$ regions, especially towards the limb. The LARS-measured center-to-limb convective blueshift for Fe I 5434 $\text{\AA}$ is highlighted in Figure \ref{fig:cbs}. 

\texttt{GRASS-E} takes a nearest-neighbor approach to determine the convective blueshift and line asymmetry at a given solar position. In other words, to create the synthetic disk-resolved line profiles, \texttt{GRASS-E} determines the granulation signatures at a given solar position based on the nearest solar axis and $\mu$ value from the input data sampling. As a result, the line bisector and convective blueshift CLV used by \texttt{GRASS-E} resembles a piecewise constant function, which is not reflective of observed CLV. Observed convective blueshift CLVs tend to resemble more of a polynomial shape as a function of $\mu$, with maximum convective blueshift occurring off-center at $\mu$ values between 0.5 and 0.9 \citep{2019A&A...624A..57L, 2023A&A...680A..62E}. During an eclipse, where partial blocking of the solar disk occurs, problems may arise from using a piecewise constant function for CLV. As the Moon crosses the solar disk during an eclipse, specific $\mu$ regions will be probed where the use of nearest-neighbor line information may be inadequate. In addition, near maximum eclipse, light is only collected from near-limb regions of which \texttt{GRASS-E} only has input observations up to $\mu$ of 0.2 and for some lines only up to $\mu$ of 0.3 (as is the case for Fe I 5434 $\text{\AA}$). Therefore, near maximum eclipse \texttt{GRASS-E}'s nearest-neighbor approach is likely unable to accurately reproduce near-limb (past the smallest observed $\mu$) line changes. This may explain the residual feature at maximum eclipse shown in Figure \ref{fig:default_models}. Therefore, when modeling a solar eclipse, it may be crucial to have center-to-limb observations at more finely sampled $\mu$ values. This is especially true near the limb, where a nearest-neighbor approach would fail to capture any extremes in the CLV, such as near-limb redshifts. 

\subsection{Optimizing Convective Blueshift}
\label{subsec:optim_CB}
As an alternative to \texttt{GRASS-E}'s granulation model, we fit a second-order polynomial for convective blueshift as a function of $\mu$. The polynomial form allows for continuous sampling of $\mu$ from 1.0 to 0.0, which is currently not captured by \texttt{GRASS-E}. 
The second-order polynomial coefficients are determined by optimizing the residual RMS of model I to the NEID observations using an ordinary least squares regression. The regression fits for the shape-dependent coefficients ($a$ and $b$) along with a constant velocity offset. Since we normalize both the NEID-observed and simulated RVs before comparing them, the constant offset cancels out and does not affect the simulated RVs. Similarly, we omit the $+c$ term in the second-order polynomial and only apply the shape-dependent coefficients, which capture the CLV shape. For each cell on the solar grid, the convective blueshift polynomial value is added to the projected velocity of the cell as a function of the cell's $\mu$ value. This convective blueshift is static with time, unlike the convective blueshift in \texttt{GRASS-E}, model III.   

Similar to model I, model II also does not account for any granulation signatures; therefore, model II, which computes CCF RVs unlike model I, can also be used to determine the second-order polynomial coefficients. Using the Nelder-Mead algorithm (an iterative downhill optimization method, \citealt{nelder}), we optimize the residual RMS of model II to the NEID observations to determine the second-order polynomial coefficients ($a$ and $b$). We find that the optimized coefficients using model II agree (within error bars) with the optimized coefficients found using model I. In addition, the residual RMS of models I-CB and II-CB remain consistent within $<0.05$ m s$^{-1}$ when using either set of optimized coefficients. This is not surprising since the only difference between models I and II is how they compute the RVs, and this does not impact the results as discussed in \S\ref{sec:Results}. We proceed with the optimized coefficients found using model I since ordinary least squares regression requires less runtime for optimizing second-order polynomials for multiple lines as done in \S\ref{sec:line_var}.

\begin{figure}
    \centering
    \includegraphics[scale=0.5]{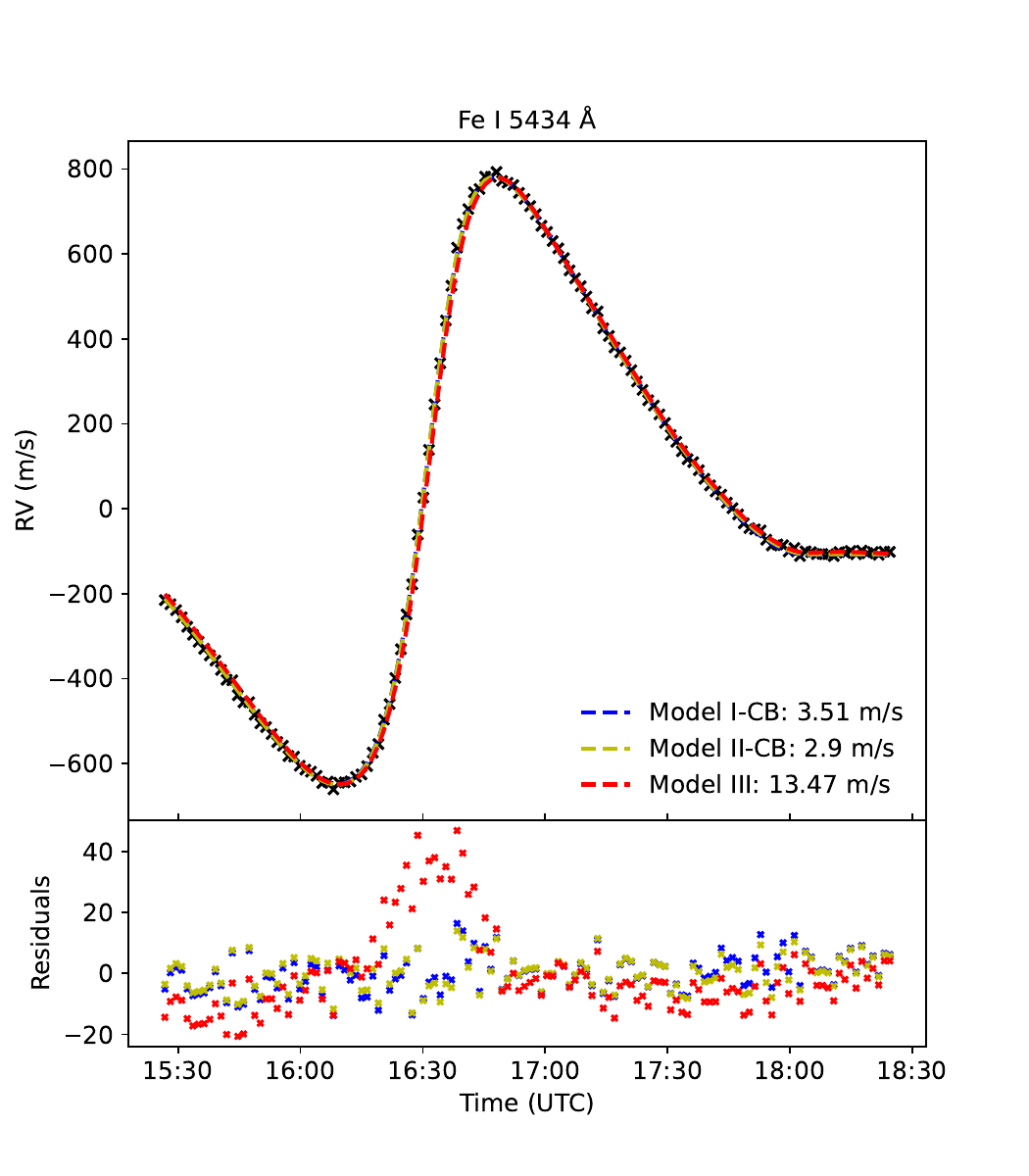}
    \caption{NEID observations of the solar eclipse RVs together with the models: models I-CB, II-CB and III. Top panel: black crosses show the NEID line-by-line RV for Fe I 5434 $\text{\AA}$. Dashed blue line shows model I-CB, average projected velocities with the optimized convective blueshift. Dashed yellow line shows model II-CB, CCF RVs from simulated spectra accounting for the optimized convective blueshift.  Dashed red line shows model III, \texttt{GRASS-E} CCF RVs with \texttt{GRASS-E} convective blueshift. All RVs are mean-subtracted. Binned RMS of each models' residuals is provided in legend. Bottom panel: residuals between the observations and each model in corresponding color. By including an optimized convective blueshift polynomial, models I-CB and II-CB now outperform GRASS-E, model III.}
    \label{fig:optim_models}
\end{figure} 

Models I and II, now models I-CB and II-CB, were rerun with the optimized polynomial for convective blueshift. Figure \ref{fig:optim_models} shows the NEID line-by-line RVs for Fe I 5434 $\text{\AA}$ along with the simulated RVs from models I-CB and II-CB with the optimized convective blueshift polynomial for Fe I 5434 $\text{\AA}$ (polynomial shown in Figure \ref{fig:cbs}). Figure \ref{fig:optim_models} highlights that models I-CB (RMS of 3.51 m s$^{-1}$) and II-CB (RMS of 2.9 m s$^{-1}$) have significant improvements in RMS compared to the original models I (RMS of 29.88 m s$^{-1}$) and II (RMS of 29.95 m s$^{-1}$) that exclude convective blueshift. In addition, models I-CB and II-CB outperform model III (RMS of 13.47 m s$^{-1}$), which uses the \texttt{GRASS-E} granulation model, by a factor of more than 3 in RV RMS. While model III utilizes an empirical granulation model, it is unable to reach the level of precision achieved by models I-CB and II-CB. This discrepancy highlights that the current \texttt{GRASS-E} granulation model does not fully capture the complex CLV required to accurately reproduce the solar eclipse observations. 
Furthermore, the inclusion of an optimized convective blueshift polynomial allows models I-CB and II-CB to reach the estimated $\sim$3 m s$^{-1}$ photon noise of the entire eclipse for the Fe I 5434 $\text{\AA}$ line. 

\subsection{Line-by-Line Variability}
\label{sec:line_var}
\begin{figure*}[!htb]
    \epsscale{1.155}
    \plotone{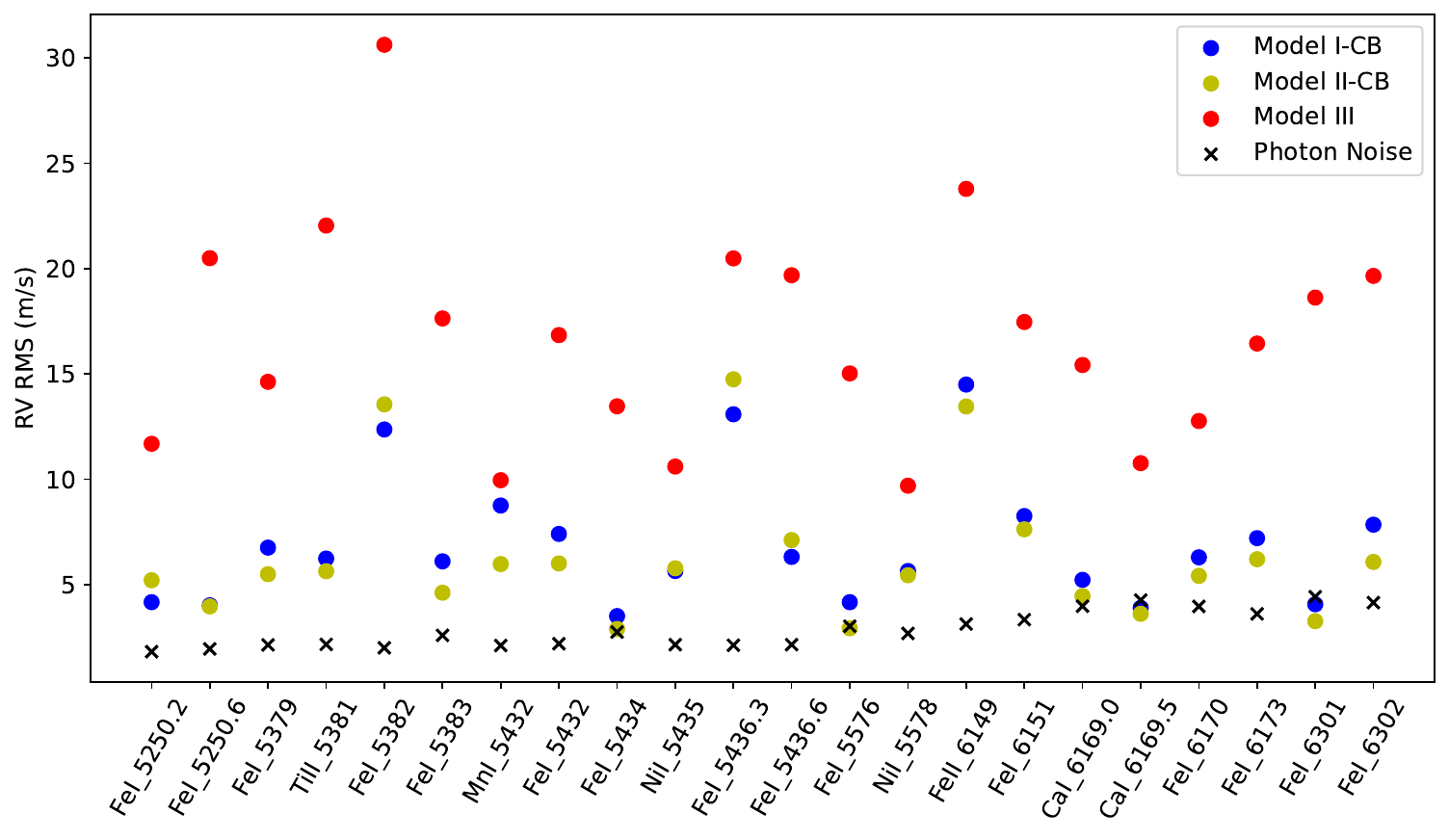}
    \caption{Eclipse RV RMS of each modeled spectral line for three models binned into 4 exposures (5.4 min) for p-modes. Model I-CB (blue): weighted average projected velocity with optimized convective blueshift; model II-CB (yellow): CCF RV from simulated spectra shifted according to optimized convective blueshift; model III (red): CCF RV from simulated \texttt{GRASS-E} granulation-driven spectra with LARS convective blueshift. Black x's depict average photon noise limit across the solar eclipse. Across all considered spectral lines, model III (\texttt{GRASS-E}) performs the weakest.}
    \label{fig:all_lines}
\end{figure*}

The Fe I 5434 $\text{\AA}$ line provides a good case study to investigate the different RV models; however, line-by-line variability is known to exist due to line varying excitation and ionization potentials, which influence the line formation location in the atmosphere. Since individual line formation probe unique ranges of heights in the solar atmosphere, different line lists will generate different measured RVs \citep{2022A&A...664A..34A, 2024A&A...683A.106A}. To investigate potential line-by-line variability, we run our models for all 22 \texttt{GRASS-E} template lines. We proceed with model I-CB that includes the optimized convective blueshift polynomial, model II-CB that also includes the optimized convective blueshift polynomial, and model III that keeps the \texttt{GRASS-E}'s convective blueshift treatment. Table \ref{tab:poly_coeffs} shows the optimized coefficients of the line-by-line convective blueshift polynomials. We only report the shape-dependent coefficients ($a$ and $b$) since the simulated RVs are insensitive to the $+c$ term due to the normalization of the RVs. We confirm that each of these optimized convective blueshift polynomials follow the expected trend of reduced convective blueshift towards the limb. Figure \ref{fig:all_lines} highlights the four-exposure binned RMS of each model and each line. Across all 22 lines, model III performs the weakest --- emphasizing that an RV model with a simple convective blueshift polynomial outperforms \texttt{GRASS-E}'s advanced empirical granulation treatment that accounts for convective blueshift and line asymmetry. For the majority of the spectral lines, model II-CB performs better than model I-CB, suggesting that once convective blueshift is accounted for, the method of RV computation can now impact the results. Yet, the majority of the models fail to reach the average photon noise limit for the considered observations (depicted in black x's). In addition, the RV RMS varies across the different spectral lines. We explore correlations between the RV RMS and various line parameters, but find no obvious correlations with any single parameter, which is consistent with the findings in \cite{2024AJ....168...46P}. The variability of model III could be driven instead by differences in the LARS data used in \texttt{GRASS-E}, such as varying $\mu$ coverage from center to limb. The combined model II-CB weighted RMS of all 22 \texttt{GRASS-E} template lines (weighted by inverse squared RMS) is $\sim$5.2 m s$^{-1}$ over the RM signal.

\begin{table}[ht]
\begin{tabular}{lccc}
\hline
Spectral Line ($\text{\AA}$) & $a$ & $b$ \\
\hline
Fe I 5250.2 & 1.625 & -2.294 \\
Fe I 5250.6 & 1.421 & -2.193 \\
Fe I 5379 & 1.519 & -2.180 \\
Ti II 5381 & 0.497 & -0.938 \\
Fe I 5382 & 2.075 & -3.088 \\
Fe I 5383 & 0.978 & -1.646 \\
Mn I 5432 & 0.642 & -0.945 \\
Fe I 5432 & 1.616 & -2.355 \\
Fe I 5434 & 0.637 & -1.221 \\
Ni I 5435 & 1.431 & -1.997 \\
Fe I 5436.3 & 1.893 & -2.697 \\
Fe I 5436.6 & 1.379 & -2.047 \\
Fe I 5576 & 0.783 & -1.356 \\
Ni I 5578 & 2.003 & -2.737 \\
Fe II 6149 & 1.442 & -2.139 \\
Fe I 6151 & 0.919 & -1.390 \\
Ca I 6169.0 & 0.638 & -1.089 \\
Ca I 6169.5 & 0.721 & -1.180 \\
Fe I 6170 & 0.653 & -1.067 \\
Fe I 6173 & 1.410 & -2.065 \\
Fe I 6301 & 0.660 & -1.230 \\
Fe I 6302 & 1.548 & -2.336 \\
\hline
\end{tabular}
\caption{Line-by-line optimized coefficients for second-order convective blueshift polynomial $f(\mu) = a \mu^2 + b \mu$ (km s$^{-1}$), where $a$ and $b$ (km s$^{-1}$) are the coefficients listed in the table. The $+c$ term is omitted since the simulated RVs are relative.}
\label{tab:poly_coeffs}
\end{table}

\subsection{The Impact of Charge-Transfer Inefficiency}
\label{sec:CTI}
During readout, charge-coupled device (CCD) detectors may experience errors when transferring recorded charge between pixels, an effect known as charge transfer inefficiency (CTI). Since CTI is dependent on the total collected flux, precision RV surveys set exposure SNR triggers for all observations of a given star \citep{2009EAS....37..247B, 2021AJ....161..130G}. Therefore, our solar eclipse observations, with 76\% reduction in total collected flux at maximum eclipse and with fixed 55 second exposures, may be sensitive to CTI uncertainties. We consider how the data-model I-CB RV deviations vary as a function of line-by-line SNR and find no trend with SNR. Even the drastic 76\% drop in total flux during the eclipse does not induce significant CTI-related uncertainties, which is not surprising given NEID's 5 cm s$^{-1}$ CTI induced RV error budget \citep{10.1117/12.2232761} and the consistency with on-sky tests during NEID commissioning (NEID Team, private communication).

\subsection{Optimizing A Non-Radial Velocity Field}
\label{sec:MF}
The fact that the majority of the spectral lines we consider do not reach their photon noise limit leads us to believe that additional sources of astrophysical variability are missing from our models. We suspect longer length-scale velocity fields, such as meridional flows \citep[e.g.,][]{2010ApJ...715..500M, 2020A&A...638A..54M} and supergranulation \citep[e.g.,][]{Rincon2018, 2019A&A...625L...6M}. 

\subsubsection{Meridional Flows}
We begin exploring additional velocity fields by incorporating to our RV models a ground-based version of the meridional flows velocity field model used in \cite{2024ApJ...973...11P} (code available on GitHub, \citealt{2024zndo..10966689P}). Meridional flows are a direct result of turbulent-driven transports of angular momentum on stars' surfaces. Related plasma flows occur symmetrically poleward at velocities up to 40 m s$^{-1}$, antisymmetric from the equator \citep{2021ApJ...916...87K}. Such velocity fields are largely ignored in RV models, despite being shown to introduce 1-2 m s$^{-1}$ RV variations \citep{2020A&A...638A..54M}. Our meridional flows model is based on \cite{2021ApJ...916...87K}, which utilize even-degree ($l = 2, l = 4$) poloidal components of vector spherical harmonic expansion. We refer the reader to Section 2 and Appendix A of \cite{2021ApJ...916...87K} for specific details on the parameterization of meridional flows. The meridional flows model is a linear combination of the $l = 2, l = 4$ poloidal components. We determine the optimized meridional flows velocity field by optimizing the coefficients of the $l = 2, l = 4$ linear combination. Such optimization is completed similarly to how we determine the optimized convective blueshift polynomial in Section \S\ref{subsec:optim_CB}, by optimizing the residual RMS of model I-CB, with the $l = 2, l = 4$ Legendre polynomials for meridional flows, to the NEID observations using an ordinary least squares regression. Much like the optimized convective blueshift, the meridional flows velocity is added to the projected velocity of the cell but as a function of cell solar latitude and longitude. The poloidal axis is fixed to the Sun's rotation axis.

The optimized non-radial velocity field is consistent with the expected meridional flows velocity field found using the model from \cite{2024ApJ...973...11P}.
When including the optimized non-radial velocity field, model I-CB's binned RMS drops from 3.51 to 3.06 m s$^{-1}$ for Fe I 5434 $\text{\AA}$.
However, this improvement manifests across all 22 \texttt{GRASS-E} template lines inconsistently, with some lines having little to no improvement and other lines having a drop in RMS up to 2 m s$^{-1}$ when including the optimized non-radial velocity field. This incoherence across the spectral lines suggests that rather than capturing the true velocity field, the optimized non-radial velocity field may provide the model with additional degrees of freedom that could absorb line-by-line variability.

\subsubsection{Supergranulation}
Supergranulation refers to the convection process associated with supergranules, which consist of clusters of individual granules and have diameters ranging from 30-35 Mm \citep{Rincon2018}. Unlike granulation that has predominantly vertical plasma flows, supergranulation occurs mostly horizontally with 300-400 m s$^{-1}$ horizontal flows and weaker 20-30 m s$^{-1}$ vertical flows \citep{Rincon2018}. 

Suspecting supergranulation,
we turn to using solar velocity fields from the Solar Dynamics Observatory’s (SDO) Helioseismic Magnetic Imager (HMI; \citealt{2012SoPh..275..207S}). Following the methods from \cite{2016MNRAS.457.3637H} and \cite{2024ApJ...973...11P}, we correct the HMI Dopplergrams for satellite motion and differential rotation in order to estimate a solar velocity field that accounts for all solar processes, such as convective blueshift, supergranulation and meridional flows. We input the SDO/HMI velocity field for 6173 $\text{\AA}$ into our model I, which computes weighted average projected RVs excluding the optimized convective blueshift polynomial, and find that the resulting RM curve is very similar to the original results of model I for 6173 $\text{\AA}$ without any convective blueshift. In other words, the HMI Dopplergrams do not provide any additional information past our initial weighted average projected RVs. \cite{2016A&A...595A..26R} also found that HMI Dopplergrams do not include information of convective blueshift  

\section{Summary}
\label{sec:Summary}
The discovery of Earth-mass exoplanets requires accurate mitigation of granulation. Although the impact of granulation on spectral lines is well understood, we currently lack the ability to accurately account for granulation within our RV models. Convective blueshift, specifically, is well-known to influence the shape of a RM curve. Therefore, we use RVs of the 2023 October 14 solar eclipse from the NEID solar feed and NEID spectrograph to test different RV models and probe our ability to model the influence of convective blueshift. We present three models: model I computes weighted average projected RVs, model II computes CCF RVs from simulated spectra with no granulation signatures, and model III computes CCF RVs from simulated \texttt{GRASS-E} granulation-driven spectra. We find that \texttt{GRASS-E} performs better than the models with no knowledge of granulation signatures. However, when optimizing and including a convective blueshift polynomial in the models that previously did not account for granulation signatures, we are able to outperform \texttt{GRASS-E}'s ability to recover the solar eclipse RM curve by more than a factor of 3 in RV RMS. In addition, we find that including a non-radial velocity field in our model allows us to further reduce our RV RMS for the majority of the considered spectral lines. 

We cannot fully dismiss sources of uncertainty such as: potential inaccuracies in the adopted limb darkening and differential extinction corrections, incomplete p-mode mitigation, and the imperfect granulation-induced CLV. Regardless, our efforts show improvements from previous efforts to model convective blueshift during a solar eclipse. For example, the 2015 March 20 solar eclipse observed by the IAG Fourier Transform Spectrograph was modeled similarly to our model I and I-CB but using a sigmoid parameterization of convective blueshift \citep{2016A&A...595A..26R}. Their version of our model I-CB (using 5000 -- 6200 and 6500 -- 6700 $\text{\AA}$) has structured residuals up to $\sim$40 m s$^{-1}$. The residual structure they find is similar in shape to the residuals from our models I and II, which peak at $\sim$75 m s$^{-1}$. When using interpolated synthetic line profiles instead to compute RVs, \cite{2016A&A...595A..26R} are able to reduce their peak residual to $\sim$25 m s$^{-1}$, but the residual structure persists. On the other hand, our models using an optimized convective blueshift polynomial have unstructured residuals, suggesting that the residual structure is dependent on the correction of convective blueshift.

We find that when attempting to model the solar eclipse RM curve, \texttt{GRASS-E} is empirically limited by the sampling of existing center-to-limb observations as \texttt{GRASS-E} currently takes a piecewise constant function approach for its CLV models. To overcome this limitation, inter-/extra-polations of the \texttt{GRASS-E} v2.0 template line asymmetries and convective blueshift could be conducted in order to better resemble continuous CLV. Such modifications to \texttt{GRASS-E} are beyond the scope of this paper, which solely aims to evaluate \texttt{GRASS-E}'s current granulation model. In addition, more finely sampled spectral observations at well-defined $\mu$ values would aid our understanding and modeling of CLV. Such observations may be obtained with the 4 m Daniel K. Inouye Solar Telescope (DKIST, \citealt{2012ASPC..463..377R}),  upcoming Paranal solar ESPRESSO Telescope (PoET, \citealt{2025arXiv250508540S}), and upcoming MIcromirrors for Resolved Adaptable Solar Observations Limb-to-limb (MiraSOL, \citealt{10.1117/12.3080954, 10.1117/12.3104198}). The advent of the 30 m class ELTs and their associated high-resolution spectrometers (e.g., The GMT-Consortium Large Earth Finder \citep{2016SPIE.9908E..22S} and ANDES \citep{2024eas..conf.2522M}) will allow us to perform similar tests on stars with transiting planets, probing granulation across cool, Sun-like stars. However, since granulation-induced CLV remains extremely challenging to measure spectroscopically, the most promising way to parameterize granulation-induced CLV may just be through highly precise RV observations of a solar eclipse with instruments such as NEID. Our work is an important step in this direction. 

\section{Acknowledgments}
We thank the anonymous referees for their thorough reviews, which have improved the clarity of this work.

AI tools were used to assist in debugging the source code presented in this work; the tool used was Claude Opus 5.0.

Based on observations at Kitt Peak National Observatory, NSF NOIRLab, managed by the Association of Universities for Research in Astronomy (AURA) under a cooperative agreement with the National Science Foundation. Data presented were obtained by the NEID spectrograph built by Penn State University and operated at the WIYN Observatory by NOIRLab, under the NN-EXPLORE partnership of the National Aeronautics and Space Administration and the National Science Foundation. Development of NEID was performed for the Jet Propulsion Laboratory, California Institute of Technology, sponsored by the United States Government under the Prime Contract 80NM0018D0004 between Caltech and NASA. We acknowledge support from NASA Grant 80NSSC26K0554. WIYN is a joint facility of NSF NOIRLab, Indiana University, the University of Wisconsin–Madison, the Pennsylvania State University, and Princeton University. Kitt Peak National Observatory sits atop I'oligam Du'ag (Manzanita Shrub Mountain). Astronomers are honored to be permitted to conduct scientific research on the sacred mountain located in the homelands of the Schuk Toak District within the Tohono O'odham Nation. We honor their past, present, and future generations, who have lived here for time immemorial and will forever call this place home. 

The Center for Exoplanets and Habitable Worlds is supported by Penn State and its Eberly College of Science. 
The authors recognize the Penn State Institute for Computational and Data Sciences (ICDS)  (RRID:SCR\_025154) for providing access to computational research infrastructure (RRID:SCR\_026424). 
This content is solely the responsibility of the authors and does not necessarily represent the views of the Institute for Computational and Data Sciences.

EG was supported by the National Science Foundation Graduate Research Fellowship Program under Grant No. DGE1255832. Any opinions, findings, and conclusions or recommendations expressed in this material are those of the authors and do not necessarily reflect the views of the National Science Foundation.

MLP was supported by the Flatiron Research Fellowship at the Flatiron Institute, a division of the Simons Foundation. 

This research has made use of NASA's Astrophysics Data System Bibliographic Services and NASA’s Navigation and Ancillary Information Facility data services.

\facilities{NEID/WIYN (NEID solar feed)}

\software{Julia \citep{Julia}, JuliaGPU \citep{JuliaGPU}, \texttt{GRASS} (\citealt{palumbo_iii_2026_23814656}), \texttt{RvSpectML} \citep{eric_ford_2021_4593963}, \texttt{Matplotlib} \citep{Hunter2007}, \texttt{barycorrpy} \citep{2018RNAAS...2....4K}, \texttt{Optim} \citep{Optim.jl-2018}, \texttt{pandas} \citep{mckinney-proc-scipy-2010}, SPICE \citep{ACTON199665}, \texttt{astropy} \citep{2013A&A...558A..33A}, \texttt{sunpy} \citep{sunpy_community2020}, \texttt{numpy} \citep{5725236}}

\bibliography{refs}
\bibliographystyle{aasjournalv7}
\end{document}